\documentclass[conference,a4paper]{APSIPA2021}
\usepackage{amsmath}
\usepackage{amsfonts}
\usepackage{graphicx}
\usepackage{multirow}
\usepackage{etoolbox,siunitx}
\usepackage{array}
\usepackage{algorithm}
\usepackage{algpseudocode}
\usepackage{threeparttable}
\usepackage[backend=biber,style=ieee,]{biblatex}
\usepackage {booktabs}
\usepackage{utfsym}
\usepackage{geometry}
\usepackage{fancyhdr}

\newcommand\sprelu{\operatorname{PReLU}}
\newcommand\layernorm{\operatorname{LN}}
\newcommand\bottleneckconv{\operatorname{BottleNeckConv}}
\newcommand\conv{\operatorname{Conv1D}}
\newcommand\tconv{\operatorname{T-Conv1D}}
\newcommand\mamba{\operatorname{Mamba}}

\fancypagestyle{firststyle}{
  \fancyhf{}
  \fancyhead[C]{2024 Asia Pacific Signal and Information Processing Association Annual Summit and Conference (APSIPA ASC)}
}

\begin{document}

\title{U-Mamba-Net: A highly efficient Mamba-based U-net style network for \\ noisy and reverberant speech separation}

\author{
\authorblockN{
Shaoxiang Dang\authorrefmark{1},
Tetsuya Matsumoto\authorrefmark{1}, 
Yoshinori Takeuchi\authorrefmark{2}, and
Hiroaki Kudo\authorrefmark{1}
}

\authorblockA{
\authorrefmark{1}
Graduate School of Informatics, Nagoya University, Nagoya, Japan}

\authorblockA{
\authorrefmark{2}
School of Informatics, Daido University, Nagoya, Japan\\ 
E-mail: dang.shaoxiang.s0@s.mail.nagoya-u.ac.jp}
}

\maketitle
\thispagestyle{firststyle}
\pagestyle{fancy}

\begin{abstract}
The topic of speech separation involves separating mixed speech with multiple overlapping speakers into several streams, with each stream containing speech from only one speaker. Many highly effective models have emerged and proliferated rapidly over time. However, the size and computational load of these models have also increased accordingly. This is a disaster for the community, as researchers need more time and computational resources to reproduce and compare existing models. In this paper, we propose U-mamba-net: a lightweight Mamba-based U-style model for speech separation in complex environments. Mamba is a state space sequence model that incorporates feature selection capabilities. U-style network is a fully convolutional neural network whose symmetric contracting and expansive paths are able to learn multi-resolution features. In our work, Mamba serves as a feature filter, alternating with U-Net. We test the proposed model on Libri2mix. The results show that U-Mamba-Net achieves improved performance with quite low computational cost.
\end{abstract}

\section{Introduction}

Speech separation in complex environments is one of the most challenging branches of the speech processing community \cite{1,1-1,6}. Noise, reverberation, and other factors severely interfere with the desired signal, making it even more difficult to capture long-term dependencies, which has long been valued by the community \cite{2-3,2-5,2-6}.

In the endeavor to capture long-term dependencies, recurrent neural networks (RNNs) are the first popular structure that demonstrates effectiveness on both time-frequency \cite{2,2-1,2-2,2-3,2-4,2-5} and time-domain processing \cite{3,3-1,3-2,5,6,13}. In time-domain separation, RNNs are typically used in conjunction with dual-path (DP) structure \cite{3}. The DP structure assists the network in modeling global information, complementing the processing of local information. Later on, self-attention mechanism like Transformer \cite{4}, as an alternative to RNNs, that allows parallel computing is introduced into the DP structure \cite{5}. As for drawbacks, it adds extra computational overhead, and the quadratic scaling of the self-attention mechanism in transformers also make training time-consuming and resource-intensive, which has been a growing consensus and concern in this community.

The cascaded multi-task learning (CMTL) methods advocates solving complex speech separation task structurally \cite{1}. Concretely, it decomposes a complex task into simpler, sequential sub-tasks. Through progressive supervision, cascaded multi-task structure can achieve further improvements in performance. However, due to the stacking of sub-modules, CMTL cannot effectively control the size of the model. Worse still, there may be gradient conflicts between modules, which significantly limits the model's capabilities \cite{6}.

In this work, we shed light on U-net, which though was developed for biomedical image segmentation \cite{7}. The U-net architecture consists entirely of convolutional neural layers, making it compact in size and low in computational complexity. Demucs \cite{8} and SuDoRMRF \cite{9} are successful attempts at using the U-Net architecture for music sound separation and audio source separation, respectively. To address its inability to learn global relationships, we add a Mamba module after each U-Net block. Mamba is a selective structured state space sequence model \cite{10}. Concretely, Mamba features high-order polynomial projection operators (HiPPO) initialization and input-dependent structure while maintain linear computational complexity. HiPPO initialization assists in decomposing signals into basis functions \cite{11}. We conduct experiments in the simulated noisy and reveberant version of open-source public dataset Libri2mix \cite{12}. According to performance comparison of proposed model and previous models, we find that U-Mamba-Net is not only able to performance better cross various metrics, but also achieve so with low computational resource.

\begin{figure}[t]
\begin{center}
\includegraphics[width=0.49\textwidth]{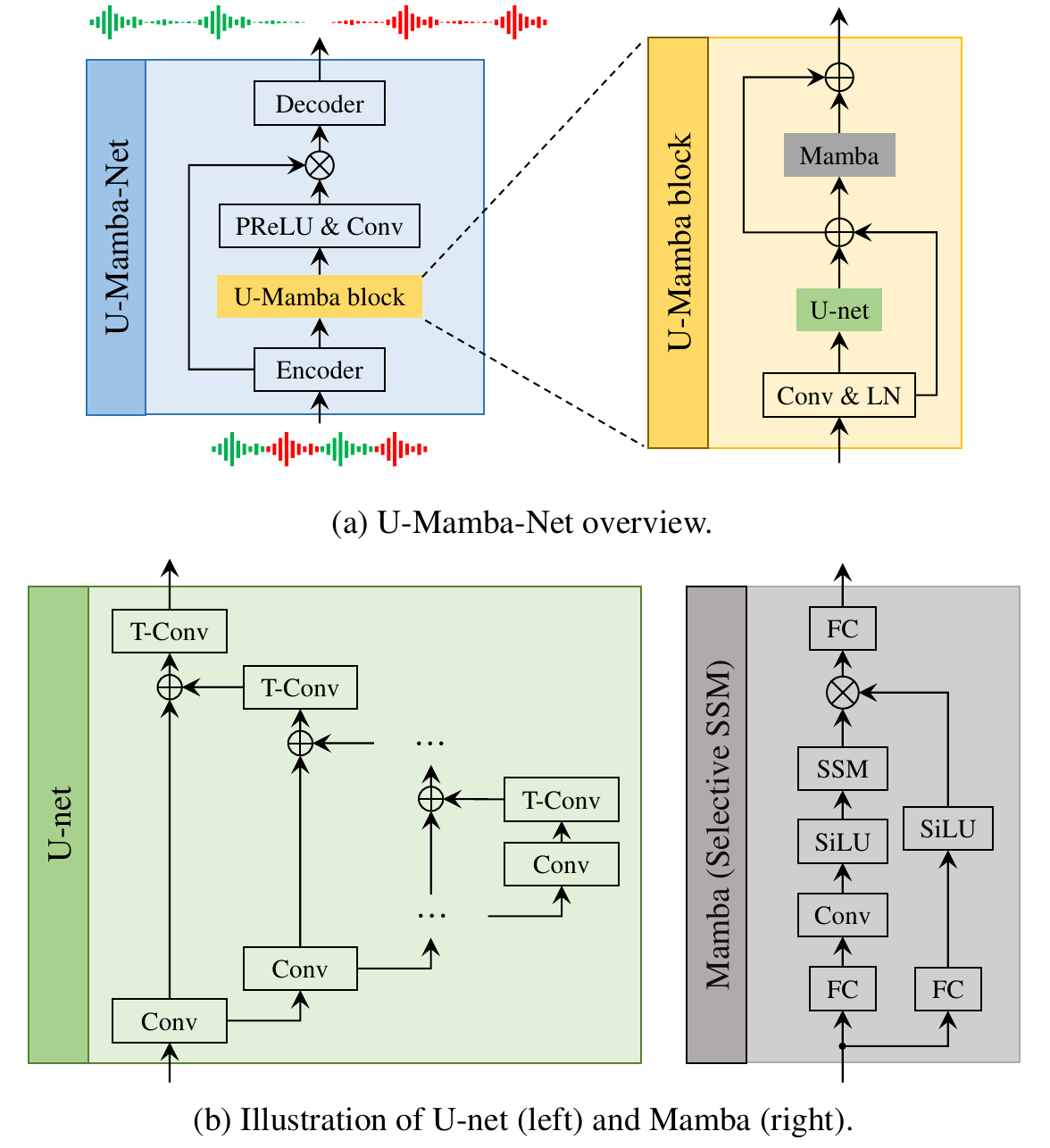}
\end{center}
\caption{Overview of U-Mamba-Net.}
\vspace*{-3pt}
\label{fig:1}
\end{figure}

\section{Proposed U-Mamba-Net}
U-Mamba-Net consists of an encoder, U-Mamba blocks, and a decoder, similar to the prevailing structures used previously \cite{13,13-1}. The encoder is a 1-dimensional convolutional layer that maps waveform to time-frequency like representation. U-Mamba blocks aim to learn features with high representational capability. A convolutional layer following U-Mamba blocks assists estimating masks for all sources. After applying the estimated masks to the representations of the mixed speech, the features of all sources are decoded by the 1-dimensional transposed convolutional layer, resulting in the final separated speeches, The overview is illustrated in Fig. \ref{fig:1}.

\subsection{U-Mamba blocks}
Given the feature $\boldsymbol{X}  \in\mathbb{R}^{F\times T}$ of a noisy and reverberant mixed speech, the goal of the $B$ stacked U-mamba blocks is to estimate robust representation $\boldsymbol{M}  \in\mathbb{R}^{F\times T}$. Then, a convolutional layer helps to generate masks equal to the number of sources $S$. As the core part of our proposed model, one U-Mamba block mainly owns a U-net module and a Mamba module. The U-net module is composed of $L$ successive downsampling and upsampling layers, with each pair of sub-layers at corresponding depths connected by residual connections. Subsequently, the output of the U-net is fed into the Mamba module. The U-Mamba block outputs the estimated representations with another residual connection to the output of the U-net in this module. The detailed statement is presented in Algorithm \ref{alg:1}.

\begin{algorithm}[t]
\caption{An algorithm for $b$-th U-Mamba block}\label{alg:cap}
\begin{algorithmic}
\Require{$\boldsymbol{M}_b$, $L$}
\Ensure{$\boldsymbol{M}_{b+1}$}

\State $\boldsymbol{D}^{(0)}$   $\gets$ $\sprelu$($\layernorm$($\bottleneckconv$($\boldsymbol{M}_{b}$)))

\Comment{Bottle Neck Convolution}
\State $ l\gets 1$

\While{$l \leq L$}   \Comment{Downsampling}
\State $\boldsymbol{D}^{(l)}$   $\gets$ $\layernorm$($\conv$($\boldsymbol{D}^{(l-1)}$))
\State $ l\gets l+1$
\EndWhile

\State $ l\gets L$
\State $\boldsymbol{U}^{(L)}$  $\gets$ $\boldsymbol{D}^{(L)}$
\While{$l \geq 1$}   \Comment{Upsampling}
\State $\boldsymbol{U}^{(l-1)}$   $\gets$ $\tconv$($\boldsymbol{U}^{(l)}$) + $\boldsymbol{D}^{(l-1)}$ 

\Comment{Residual connection}
\State $ l\gets l-1$
\EndWhile

\State $\boldsymbol{M}_b$   $\gets$ $\sprelu$($\boldsymbol{U}^{(0)}$)

\State $\boldsymbol{M}_{b+1}$   $\gets$ $\mamba$($\boldsymbol{M}_b$) + $\boldsymbol{M}_b$

\Comment{Mamba module and residual connection}

\State \Return $\boldsymbol{M}_{b+1}$
\end{algorithmic}
\label{alg:1}
\end{algorithm}

\subsection{Mamba}

Mamba is an extension of state space models (SSMs) \cite{14}. SSMs formulates a mapping from $\boldsymbol{x}(t) \in\mathbb{R}^{F}$ to $\boldsymbol{y}(t) \in\mathbb{R}^{F}$ via hidden state space $\boldsymbol{h}(t)\in\mathbb{R}^{N}$:

\begin{align}
\boldsymbol{h}^{'}(t) &=  \boldsymbol{A} \boldsymbol{h}(t) + \boldsymbol{B}\boldsymbol{x}(t)   \\
\boldsymbol{y}(t) &=  \boldsymbol{C} \boldsymbol{h}(t) + \boldsymbol{D} \boldsymbol{x}(t)  
\label{eqa:2}
\end{align}
where $\boldsymbol{A} \in\mathbb{R}^{N\times N}$, $\boldsymbol{B} \in\mathbb{R}^{N\times F}$, $\boldsymbol{C} \in\mathbb{R}^{F\times N}$, and $\boldsymbol{D} \in\mathbb{R}^{F\times F}$ are state matrices. $N$ is number of state space dimension.

Based on the continuous form above, its discrete form can be formulated by:
\begin{align}
\boldsymbol{h}_{t} &=  \boldsymbol{\overline{A}} \boldsymbol{h}_{t-1} + \boldsymbol{\overline{B}} \boldsymbol{x}_t   \\
\boldsymbol{y}_{t} &=  \boldsymbol{\overline{C}} \boldsymbol{h}_{t} + \boldsymbol{\overline{D}} \boldsymbol{x}_t  
\label{eqa:2}
\end{align}
where $\boldsymbol{\overline{A}}$ and $\boldsymbol{\overline{B}}$ are discretized parameters, which are converted from $\boldsymbol{A}$, $\boldsymbol{B}$, and a step parameter $\Delta$ using bilinear method:

\begin{align}
\boldsymbol{\overline{A}} &= ( \boldsymbol{I} - \frac{\Delta}{2}\boldsymbol{A})^{-1} ( \boldsymbol{I} + \frac{\Delta}{2}\boldsymbol{A}) \\
\boldsymbol{\overline{B}} &= ( \boldsymbol{I} - \frac{\Delta}{2}\boldsymbol{A})^{-1}  \Delta \boldsymbol{B} \\
\boldsymbol{\overline{C}} &= \boldsymbol{C}\\
\boldsymbol{\overline{D}} &= \boldsymbol{D}
\label{eqa:2}
\end{align}

System in equations 3 and 4 can be rewritten as a convolutional form of structured kernel $\boldsymbol{\overline{K}}$ and input $\boldsymbol{x}$.

\begin{equation}
\boldsymbol{y} = \boldsymbol{\overline{K}} \ast \boldsymbol{x},
\boldsymbol{\overline{K}} = (\boldsymbol{\overline{C}}\boldsymbol{\overline{B}},\boldsymbol{\overline{C}}\boldsymbol{\overline{A}}\boldsymbol{\overline{B}},\cdots,\boldsymbol{\overline{C}}\boldsymbol{\overline{A}}^{T-1}\boldsymbol{\overline{B}})
\label{eqa:2}
\end{equation}

Structured SSMs (S4) feature a HiPPO initialization instead of random initialization, which is confirmed to easier decompose representation into orthogonal polynomials.

Mamba improves S4 with a selective mechanism \cite{10}. Concretely, it adds input-dependency into SSM matrices via parallel scan algorithm, Furthermore, state variables are kept in SRAM, which holds a fast and efficient GPU memory hierarchy. Mamba is illustrated in the right part of Fig. \ref{fig:1}b.

\section{Experiments}

\subsection{Datasets}
We use open-source mixture dataset Libri2mix to conduct experiments \cite{12}. Source audios are from Librispeech, ambient noises are sampled from WHAM! \cite{15}. As for simulation of indoor surrounding, we exploit Pyroomacoustics toolkit to generate reverberant version of clean source \cite{16}. The simulated mixture is produced by adding each reverberant source and ambient noise together in time domain. The configuration of indoor reverebration is shown in Table \ref{tbl:1}. T60 represents the reverberation intensity, indicating the time required for the sound energy to decay by 60 dB. As a result, our training, validation, and test datasets owns 13900, 3000, 3000 samples, respectively. All the samples are in 8 kHz.

\begin{table}[t]
\begin{center}
\caption{Reverberation configuration.}
\setlength{\tabcolsep}{2mm}{
\begin{tabular}{c|c|c}
\hline
 \multirow{3}{*}{Room}& L (m) & $\mathcal{U}$(5,10) \\
             &   W (m) & $\mathcal{U}$(5,10) \\
             &   H (m) & $\mathcal{U}$(3,4) \\
\hline
        T60 & T (s) & $\mathcal{U}$(0.2,0.6) \\
\hline
        \multirow{3}{*}{Receiver}& L (m) & $\frac{L_{room}}{2} + \mathcal{U}$($-$0.2,0.2) \\
                               & W (m) & $\frac{W_{room}}{2} + \mathcal{U}$($-$0.2,0.2) \\
                               & H (m) & $\mathcal{U}$(0.9,1.8) \\
\hline
        \multirow{3}{*}{Sources}& H (m) & $\mathcal{U}$(0.9,1.8) \\
                               & Dist. (m) & $\mathcal{U}$(0.66,2) \\
                               & $\theta$ & $\mathcal{U}$(0,2$\pi$) \\
\hline
\end{tabular}}
\label{tbl:1}
\end{center}
\end{table}

\subsection{Network}
The basic hyperparameters that we use are exhibited in Table \ref{tbl:2}. For the ablation studies, we indicate different parameter while keeping the same alphabet symbols. Additionally, we evaluate several different upsampling methods. Besides T-Conv1D, we also investigate the nearest neighbor (NN) and linear upsampling techniques.

\begin{table}[t]
\begin{center}
\caption{Hyperparameters of U-Mamba-Net.}
\setlength{\tabcolsep}{1.2mm}{
\begin{tabular}{c|c|c|c}
\hline
 Encoder / Decoder & channel / window / hop   & $F$ / - / - &  128 / 41 / 20 \\
 U-net  & Input / output channel  & - &  128 \\
 Mamba & Input / output channel  & - &  128 \\
 U-net & Down / upsampling depths  & $L$ & 4 \\
 U-Mamba  & Number of repeat core  & $B$ &  16 \\
 Sources  & -  & $S$ & 2\\
 \hline
\end{tabular}}
\label{tbl:2}
\end{center}
\end{table}

\subsection{Training phase}
We train the proposed model on GeForce 4070Ti Super with a batch size of 4 and initial learning rate of 0.00015. Each sample is randomly cut into 3-s long. Notably, we report the computational load based on the input utterance with this length. We set the maximum epoch at 120. As for objective function, we adopt permutation-invariant scale-invariant single-to-noise ratio (SI-SNR) \cite{13}, shown as follows:
\begin{align}
{\mathcal L} &= -\max_{\pi \in \mathcal{P}}\frac{1}{I}\sum_i{\rm SI\text{-}SNR}(\hat{s}_{\pi(i)},s_i) 
\end{align}
where $\pi$ is the best permutation mapping set that allow overall SI-SNR to achieve maximum \cite{17}.

\subsection{Evaluation phase}
SI-SNR improvement (SI-SNRi) is an extension of SI-SNR, it checks how much information get excluded from mixture, we use SI-SNR improvement (SI-SNRi). We use signal-to-interference ratio improvement (SIRi) for separation effectiveness \cite{18}. At the perceptual aspect, we use short-time objective intelligibility (STOI) \cite{18} and perceptual evaluation of speech quality (PESQ) \cite{20}. For denoising performance, we use predicted rating of speech distortion (CSIG), predicted rating of background distortion (CBAK), and predicted rating of overall quality (COVL) \cite{21}. Additionally, we present the model size and computational load in terms of Giga Multiply-Add Operations per Second (GMACs) \cite{22}.

\section{Experimental results}

\subsection{Main results}
We compare the results of U-Mamba-Net with several previous models in Table \ref{tbl:3}. In the implementation of CMTL using DPRNN, we adopt an Enhancement Priority Pipeline (EPP). We can observe that, firstly, the proposed U-mamba-net achieves the best results cross major metrics. It is 0.92 dB better than DPRNN in SI-SNRi. Compared to CMTL, which uses DPRNN as sub-modules, U-mamba-net is still 0.42dB higher. Moreover, U-Mamba-Net model size is 20\% smaller than DPRNN (CMTL). Most importantly, its computational efficiency (measured in GMACs) is only one-sixteenth of that of DPRNN (CMTL) (2.5 vs. 40.2) and one-ninth of that of DRRNN (2.5 vs. 23.9). This strongly demonstrates the proposed model's high efficiency and effectiveness.

\begin{table}[t]
\begin{center}
\caption{Main results.}
\setlength{\tabcolsep}{0.75mm}{
\begin{tabular}{lccc r@{.}l r@{.}l r@{.}l }
\hline
 Methods  & E2E or CMTL & SI-SNRi & SDRi  & \multicolumn{2}{c}{SIRi}  &\multicolumn{2}{c}{\#Param (M)} & \multicolumn{2}{c}{GMACs}\\
\hline
TasNet & E2E   & 5.70& 5.05& 10&84 & 23&2 M & 27&8\\
SuDoRM-RF  & E2E    & 2.90&  3.36 & 6&14 & 2&6 M & 3&6\\
SuDoRM-RF+ & E2E   & 5.33& 6.05& 11&02 & 2&7 M & 3&0\\
Conv-TasNet & E2E    & 6.88 &7.17 & 14&32 & 6&3 M & 18&7\\
DPRNN & E2E   & 7.59& 7.88 & 15&16 & 3&7 M & 23&9\\
DPRNN & CMTL  & 8.08&  8.62 & 16 &39 & 5&6 M & 40&2\\
\hline
U-Mamba-Net & E2E   &  \bf{8.50}&  \bf{8.62} &  \bf{17} & \bf{67}  & \bf{4}&\bf{4 M} & \bf{2} & \bf{5}\\
\hline

\end{tabular}}
\label{tbl:3}
\end{center}
\end{table}



\subsection{Performance of perceptual quality and denoising effectiveness}

\begin{table}[h]
\begin{center}
\caption{Perceptual and denoising performance.}
\setlength{\tabcolsep}{0.75mm}{
\begin{tabular}{lc|cc|ccc}
\hline
\multirow{2}{*}{Methods}& \multirow{2}{*}{E2E or CMTL} & \multicolumn{2}{c|}{Perceptual} & \multicolumn{3}{c}{Denoising} \\
  &  & STOI & PESQ  &  CSIG & CBAK & COVL\\
\hline
SuDoRM-RF+  & E2E  & 65.92 &1.52 & 2.37& 1.61 & 1.84\\
Conv-TasNet  & E2E   & 69.61 & 1.60& 2.58 & 1.76  & 2.00\\
DPRNN  & E2E  & 71.86& 1.67 & 2.65 & 1.83 & 2.07\\
DPRNN&CMTL & 73.19 & \bfseries 1.75 &  \bfseries2.75 &  \bfseries2.16 &  \bfseries2.18\\
\hline
U-Mamba-Net & E2E   & \bfseries 73.99 & 1.70 &  2.56  & 1.89 & 2.05\\
\hline

\end{tabular}}
\label{tbl:4}
\end{center}
\end{table}

\begin{table*}[t]
\begin{center}
\caption{Ablation studies.}
\setlength{\tabcolsep}{1mm}{
\begin{tabular}{ccccccccccccccc}
\hline
$F$&  $R$   & $L$&  Upsampling &SI-SNR & SI-SNRi & SDRi  & SIRi & STOI (\%) & PESQ & CSIG & CBAK & COVL& \#Param (M) & GMACs\\
\hline
64 &16 &4& T-Conv1D & 1.29 & 7.12 & 7.59 & 14.64 & 70.59 & 1.60 & 2.52 & 1.75 & 1.97 & 1.3 M & 0.7\\
128 &12 &4& T-Conv1D & 2.31 & 8.14 & 8.39 & 16.31 & 73.28 & 1.68 & 2.53 & 1.85 & 2.02 & 3.3 M & 1.9 \\
128 &16  &4& T-Conv1D & 2.67 & 8.50 & 8.62 &17.67  &73.97  & 1.70 & 2.53 & 1.89 & 2.04 & 4.4 M & 2.5\\
128 &20 &4 & T-Conv1D & 2.76 & 8.59& 8.76 & 17.46 & 74.25 & 1.71 & 2.56 & 1.90 & 2.06 & 5.5 M & 3.1\\
128 &16  &8& T-Conv1D & 2.59 & 8.42 & 8.52 & 17.39  & 73.72 & 1.69 & 2.50 & 1.87 & 2.01 & 4.6 M & 2.5\\
128 &16  &4& NN & 2.70 & 8.52& 8.66 &  17.44 & 73.99 & 1.70 & 2.56 & 1.89 & 2.05 & 4.4 M & 2.5\\
128 &16  &4& Linear & 2.70 & 8.53 & 8.70 & 17.54  & 74.06 & 1.71 & 2.55 &1.89  & 2.05 & 4.4 M & 2.5\\
192 &16  &4& T-Conv1D & 3.02 & 8.85&  8.90& 18.15  & 74.87 & 1.74 & 2.64 & 1.93 & 2.11 & 9.7 M & 5.3 \\
\hline
\end{tabular}}
\label{tbl:5}
\end{center}
\end{table*}

In this section, we compare the U-Mamba-Net with previous models in terms of perceptual quality and denoising performance in Table \ref{tbl:4}. In terms of the STOI metric, U-Mamba-Net consistently maintains its advantage, just as the results and conclusions in the previous section. It also shows superiority in PESQ, CSIG, CBAK, and COVL when compared to all models except DPRNN (CMTL). However, DPRNN (CMTL) outperforms U-Mamba-Net in these metrics, particularly in CBAK. This indicates that a single-task model's performance on specific sub-task is not as effective as that of a CMTL architecture incorporating additional processing modules. We believe the reason is that DPRNN (CMTL) can leverage additional information like noise-free mixture as intermediate label to supervised model in a step-by-step manner. This suggests that combining specialized modules in a multi-task framework can offer superior performance by addressing various aspects of the problem more comprehensively.

\subsection{Ablation studies}

To better understand the role of each parameter in the model, we design several ablation experiments.
Their results are displayed in Table \ref{tbl:5}. Firstly, the impact of feature dimensions on the model is most noticeable. As the feature dimension $F$ increases from 64 to 128 to 192, the separation performance significantly improves like in SI-SNRi from 7.12 dB to 8.50 dB to 8.85 dB, but the model size also increases accordingly. Increasing the number of U-mamba-Net blocks $R$ also enhances the model's performance, although the improvement is not as pronounced. With 20 blocks, the model reaches an SI-SNRi score of 8.59 dB. Increasing the depth $L$ of the model has a negative effect on the test set. We speculate that for separation tasks, excessively low resolution may not be beneficial. Using NN and linear upsampling method though has a subtle improvement on the test set, we observe a decreased performance in the validation set during training. Therefore, upsampling methods are less critical than any other hyperparameters else.

\subsection{Visualization}

\begin{figure}[t]
\begin{center}
\includegraphics[width=0.46\textwidth]{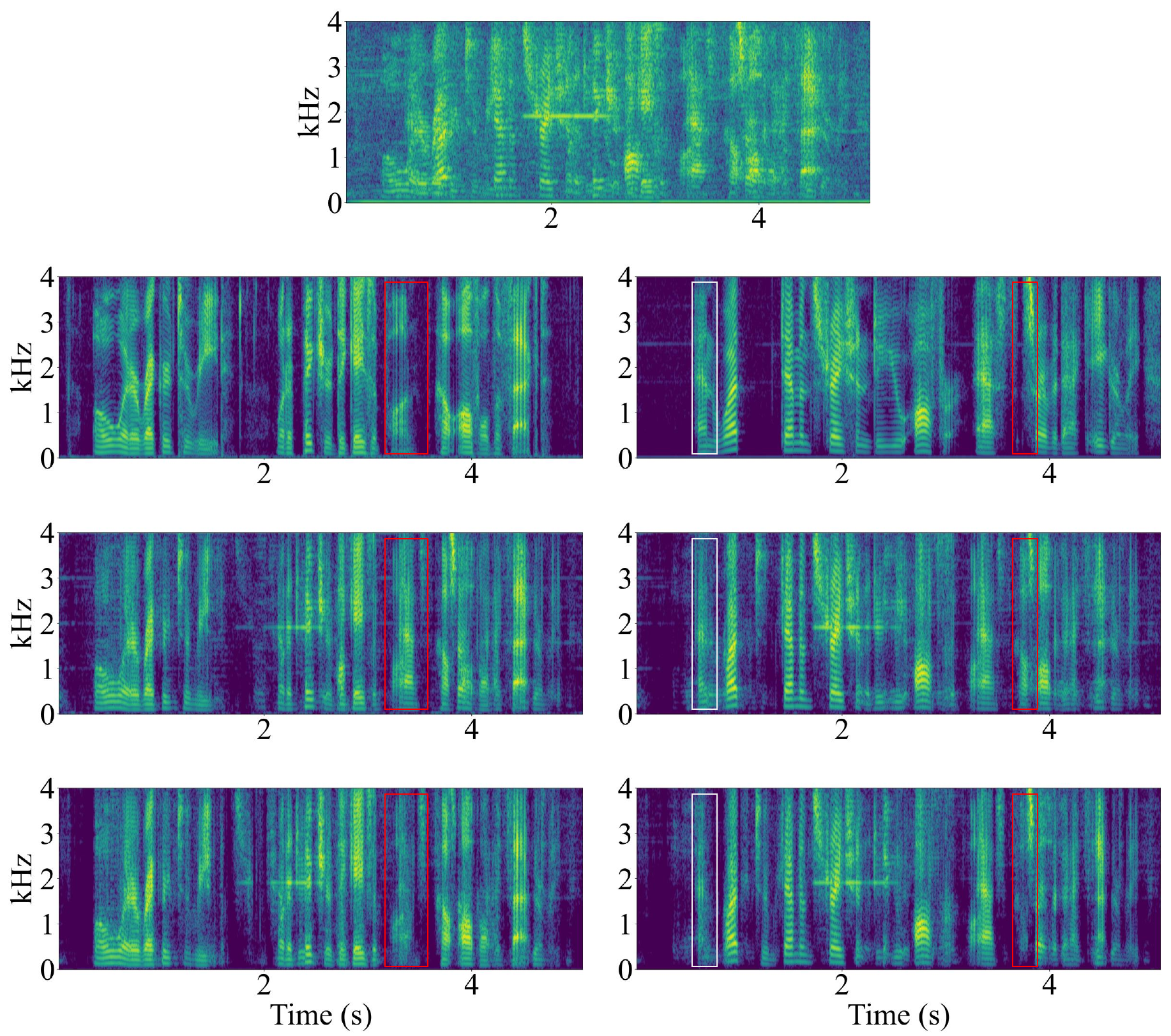}
\end{center}
\caption{Spectrogram of separation results. The sole spectrogram in the first row is noise and reverberant mixture. The following two spectrograms in the second row are ground truths. Third row are displaying the spectrograms of two separated results by DPRNN model. The last two are estimation of U-Mamba-Net. The red boxes highlight the places where DPRNN makes wrong separation, but U-Mamba does not. The white box outlines the place where U-Mamba-Net performs worse. Because the fundamental frequencies and harmonics of U-Mamba-Net are not as clear as those of the DPRNN.}

\vspace*{-3pt}
\label{fig:2}
\end{figure}

This section visualizes an example of separation results using DPRNN and U-Mamba-Net in Fig \ref{fig:2}. First, from the spectrograms of the results, both DPRNN and U-Mamba-Net exhibit effective denoising and dereverberation, as evidenced by the clarity of the spectrograms of separations. Regarding the separation results, DPRNN exhibits more erroneous separations, where information originally belonging to one source is allocated to another source. This issue is highlighted by the red boxes in the figure. We believe this problem arises from the model's less robust modeling of long-term dependencies. U-Mamba-Net also has some weaknesses. For instance, as shown in the white boxes, the generated speech spectrogram lines are not as clear as those produced by DPRNN. Even worse, when we listen to the separated speech, its perceptual quality remains relatively low, which is consistent with the STOI score indicated in the previous sections.

\section{Conclusions}

In this work, we proposed a lightweight U-Mamba-Net for noisy and reverberant speech separation. U-Mamba-Net not only demonstrates impressive separation capabilities but also maintains low computational load. In terms of SNR category metrics, it surpasses many previous models, demonstrating its overall robust separation capability. However, it still shows significant gaps in denoising and other specific metrics compared to CMTL.

\section*{Acknowledgment}
This work is supported by a fellowship of the Nagoya University CIBoG WISE program from MEXT.

\printbibliography

\end{document}